\documentclass{iopjournal}
\usepackage{amsmath}
\usepackage{amssymb}
\usepackage{subcaption}
\usepackage{tikz}
\usepackage{bm}
\usetikzlibrary{shapes.misc}
\usepackage[style=numeric-comp,sorting=none]{biblatex}
\begin{document}

\articletype{Article type} %	 e.g. Paper, Letter, Topical Review...

%\title{Feasibility of non-conventional superconducting materials in multilayer structures towards SRF applications}
\title{Analytical evaluation of surface barrier and resistance in iron-based superconducting multilayers for Superconducting Radio-Frequency applications}
% several iterations with ChatGPT to find a statistically favored title by SUST's referees
% I am not sure if non-conventional should be that stressed... because we also discuss more conventional materials...

\author{Carlos Redondo Herrero$^{1}$\orcid{0009-0000-4769-9101}, and Akira Miyazaki$^{1}$\orcid{0000-0002-7232-128X}}
%%%% unlike particle physics, this community is not in alphabetical order...supervisor must be at the end

\affil{$^1$Université Paris-Saclay, CNRS/IN2P3, IJCLab, Orsay, France}

\email{carlos.redondo-herrero@ijclab.in2p3.fr}
% I suggest that you be the corresponding author for educational purpose


\begin{abstract}
New superconducting materials, particularly iron-based superconductors (IBS), have recently attracted attention for their potential applications in particle detectors and accelerators.
This paper discusses the application of these materials in multilayer structures for radio-frequency resonators used to accelerate charged particles, with the aim of improving performance compared to bulk niobium.
These materials are compared with previously studied multilayers composed of conventional superconductors in terms of the maximum magnetic field they can withstand, their surface resistance, and their power loss per unit surface area.
Finally, perspectives and future applications aimed at increasing operating temperatures are discussed.
\end{abstract}

\section{Introduction}
Superconducting radio-frequency (SRF) cavities are among the most powerful tools for confining strong high-frequency electromagnetic waves in a wide range of applications~\cite{Grav-waves-detection,Qubit-SRF,romanenko2023newexclusionlimitdark,giaccone2022designaxionaxiondark,Valente-Feliciano:accelerator}.
One of the primary goals of particle accelerator physics is to construct high-energy, high-luminosity machines in a cost-effective manner. To achieve this, SRF technology must be further studied in order to develop cavities with high accelerating fields and low surface resistance.

State-of-the-art SRF cavities are predominantly fabricated from bulk niobium (Nb), a material that becomes superconducting below $9.2$~K and exhibits a low surface resistance at $2$~K, $R_{\text{s},\text{Nb}}(T=2\text{K}) < 10\, \text{n}\Omega$, when its coherence length $\xi$ is optimized.
The superheating field of Nb is the highest among pure elemental superconductors, $B_{\text{sh},\text{Nb}} = 180\, \text{mT}$.
These properties make Nb cavities a cornerstone of particle accelerator technology, complementing copper resonators, whose $R_\text{s}$ is more than five orders of magnitude higher than that of superconducting Nb at around 1~GHz.
Nb cavities therefore represent an excellent technology for realizing long-pulsed or continuous wave (CW) accelerators, such as high-energy storage rings, high-intensity free-electron lasers, neutron sources, and heavy-ion accelerators.
Over several decades, extensive studies have focused on the engineering aspects of Nb cavities, which are now approaching their fundamental performance limits, while some open questions remain in non-equilibrium superconductivity~\cite{Q-Slope-Checchin,Martinello:2018kct}.
Another research direction aims to eliminate the use of liquid helium by increasing the operating temperature beyond 2~K, or even above 4~K, which is the typical operating temperature of Nb cavities in particle accelerators.
Therefore, the investigation of non-Nb materials is strongly motivated for further improvements in SRF cavity performance.

Various alloys (NbTi and Nb$_{3}$Sn) and high-$T_\text{c}$ superconductors (HTS), such as cuprates and pnictides, have been investigated for applications in superconducting magnets, including the dipole magnets of the Large Hadron Collider at CERN and its future successors~\cite{FCC:1,FCC:2,FCC:3}, solenoid magnets of the ITER fusion reactor~\cite{ITER:Magnets}, as well as industrial and medical applications.
Unlike direct current (DC) applications in such magnets, superconducting cavities are exposed to radio-frequency (RF), high-frequency alternating current (AC), thereby revealing a more fundamental aspect of superconducting materials: the symmetry of the superconducting gap.
Conventional superconducting alloys are $s$-wave superconductors; thus, thermal excitation of quasiparticles is suppressed by the superconducting gap $\Delta$, leading to an exponential temperature ($T$) dependence of the surface resistance~\cite{mattis58}.
\begin{equation}
    R_\text{s}\propto \sigma_1\propto\sigma_{0} e^{-\Delta/k_\text{B} T}, \label{eq:BCS-MB-simple}
\end{equation}
where $\sigma_1$ denotes the real part of the optical conductivity in the superconducting state, $\sigma_{0}$ is the conductivity in the normal conducting phase, and $k_B$ is the Boltzmann constant.
Equation~\eqref{eq:BCS-MB-simple} indicates that materials with a larger $\Delta$ than Nb can achieve lower surface resistance at the same $T$, or alternatively, can be operated at higher $T$.
The most successful alloy for SRF cavities to date is a thin (a few $\mu$m thick) Nb$_3$Sn film formed on a bulk Nb substrate via Sn vapor deposition~\cite{Posen_2017,Ito:vapor-deposition}. So far, Nb$_3$Sn has achieved the same surface resistance at $4$~K as that of bulk niobium at $2$~K.
A key challenge for Nb$_3$Sn is its mechanical brittleness, which limits this technology to small-scale prototypes~\cite{Devred:2001hc}.
Fundamental research is ongoing using various deposition techniques on alternative substrates, such as copper~\cite{GIROIRE201753:Copper}.
Another alloy, NbN, has also been extensively studied in the context of multilayer structures, as will be discussed in this paper~\cite{Kubo_2014,katayama:KEK-NbN}.

Iron-based superconductors (IBS) were introduced for SRF applications~\cite{Gurevich_2017} because even the smaller of their two superconducting gaps is larger than that of Nb, and the commonly accepted $s^{\pm}$-wave pairing mechanism is fully gapped without nodes in momentum space, thereby justifying the use of Eq.~\eqref{eq:BCS-MB-simple} as a first-order approximation~\cite{10601306}.
In addition, IBS exhibits a smaller $\sigma_{0}$, which further reduces the surface resistance, as indicated by Eq.~\eqref{eq:BCS-MB-simple}.
Moreover, IBS possesses metallic mechanical properties, unlike other HTS materials or Nb$_3$Sn, implying significant potential for large-scale applications.
One of the challenges associated with pnictides is the handling of arsenide (As), which can be topologically encapsulated in wire geometries; however, safe fabrication methods for resonators have not yet been established.
Thus, although superconducting wires based on pnictides have already been successfully fabricated for magnet applications~\cite{Ma_2012}, their development for SRF applications remains limited to FeSe or FeSeTe, which have relatively low critical temperatures~\cite{Pompeo_2020,CAS_FeSe}.

Copper-oxide-based superconductors (cuprates) represent another promising class of high-temperature superconductors (HTS) for various applications.
Despite their high critical temperatures, their use in SRF cavities remains subject to important limitations. Since cuprates are $d$-wave superconductors, the presence of gapless modes prevents efficient suppression of thermally excited quasiparticles; consequently, their surface resistance is given by~\cite{10601306}
\begin{equation}
R_{\text{s},\text{Cupr}} = aT^b + c, \label{eq:R_s_cuprate}
\end{equation}
where $a$, $b$, and $c$ are material-dependent constants. The absence of exponential suppression leads to substantially higher surface resistance compared to $s$-wave superconductors.
It has recently been shown that the surface resistance of REBCO-coated cavities is an order of magnitude lower than that of copper cavities under a static magnetic field~\cite{Krkotic:REBCO}.
Although such relatively high-loss cuprates cannot serve as an alternative to bulk Nb cavities for long-pulsed or continuous-wave accelerators, they have attracted attention for improving copper cavities in dark matter axion searches under strong magnetic fields~\cite{Golm:2784909}, as well as for short-pulse operation of accelerating cavities~\cite{dhar2025phasetransitiondynamicsinduced}.

A comprehensive study based on simplified model calculations for \textit{bulk} non-conventional superconductors was proposed by one of the authors~\cite{10601306}.
This study demonstrated that the real part of the complex optical conductivity, $\sigma_1$, of pnictides can be significantly improved compared to that of conventional superconductors, including Nb$_3$Sn.
However, the relatively large penetration depth $\lambda$ of pnictides substantially increases the volume of material contributing to Joule heating, as indicated by $R_\text{s} \propto \sigma_{1}\lambda^3$.
The analysis predicted that the RF losses would not surpass those of niobium below 4~K, even under idealized conditions.
Therefore, the potential of IBS is primarily limited to \textit{thin-film} applications and/or operation at higher temperatures.

Thin-film SRF cavities have been studied by various institutions, with the most mature approach being Nb films (a few $\mu$m thick) sputtered onto copper substrates.
Several accelerators have been constructed using this technology~\cite{LEP:SRF, LHC:SRF, ALPI:SRF, SOLEIL:SRF, ISOLDE:SRF}.
The most successful Nb$_3$Sn implementation also relies on thin-film SRF cavities.
However, thin-film cavities generally suffer from multiple issues, including nonlinear surface resistance (the so-called Q-slope problem~\cite{benvenuti99, PhysRevAccelBeams.22.073101}) and limited accelerating gradients, partly due to surface defects~\cite{Weingarten:Defects}.
Given that the inner surface area of accelerating cavities can be as large as 1~m$^2$, and considering the complexity of the film deposition process, achieving defect-free surfaces in thin-film cavities remains a significant challenge.
Therefore, fundamental and engineering solutions are required to effectively prevent RF flux penetration through such, to some extent, unavoidable defects.

Multilayer SRF cavities were proposed~\cite{GurevichMultilayer/2006} to enhance the quench field by manipulating the Bean-Livingston barrier~\cite{Bean:1964zz,Gurevich/Bean-Livingston} using multiple thin films with thicknesses between the coherence length $\xi$ and the penetration depth $\lambda$.
Enhancement of the critical fields has recently been demonstrated under applied DC fields~\cite{katayama:KEK-NbN}, whereas its realization under RF conditions remains one of the major challenges in the SRF community.
Previous studies have primarily focused on the quench fields of conventional superconducting multilayers.
In this paper, we apply the multilayer theory to IBS multilayer structures and compare the results with conventional materials.
The theory is further extended to predict the surface resistance alongside the surface barrier calculations, allowing us to propose layer parameters that simultaneously optimize power loss and enhance the quench field.

\section{Theory} \label{sec:theory}
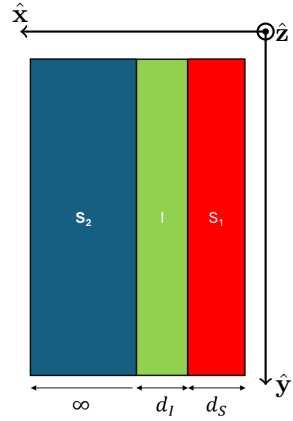
\begin{figure}[h!]
    \centering
    \begin{tikzpicture}
\tikzset{cross/.style={cross out, draw=black, minimum size=2*(#1-\pgflinewidth), inner sep=0pt, outer sep=0pt},
%default radius will be 1pt. 
cross/.default={1pt}}
    \node[anchor=south west,inner sep=0] (image) at (0,0) {\includegraphics[width=0.2\linewidth]{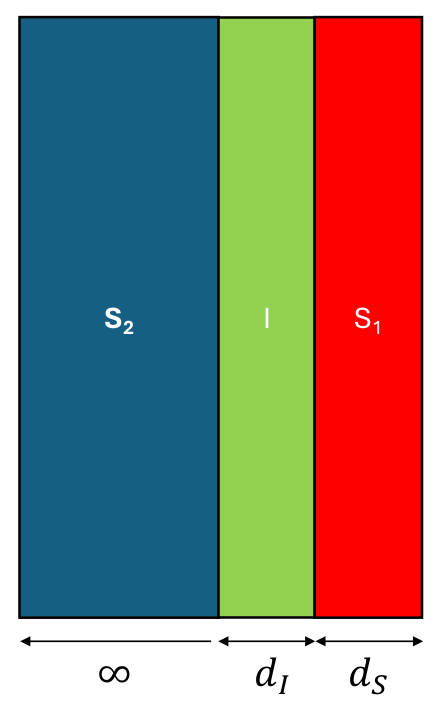}};
    \begin{scope}[x={(image.south east)},y={(image.north west)}]
    \draw[->,thick] (1.05,1.05) to (0,1.05) node[anchor=south]{{$\hat{\bm{x}}$}};
    \draw[->,thick] (1.05,1.05) to (1.05,0.1) node[anchor=west]{{$\hat{\bm{y}}$}};
    \draw[thick] (1.05,1.05) circle (3pt) node[anchor=west]{{$\hat{\bm{z}}$}};
    \filldraw[black] (1.05,1.05) circle (1pt);
    \end{scope}
    \end{tikzpicture}
    \caption{Multilayer structure, where the superconductor thin-film is denoted as $S_1$ in the red layer, the insulator layer is denoted as $I$ in the green layer, and the superconductor substrate is denoted as $S_2$ in the turquoise layer. (Color online)}
    \label{fig:ML-structure}
\end{figure}

Firstly, we introduce the multilayer theory as described in the literature~\cite{Gurevich/Bean-Livingston,Kubo_2014,Kubo_2017}, providing a concise derivation in Appendices~\ref{sec:App_EM_derivation} and~\ref{sec:App-Vortex_Calc}.
We consider a simple multilayer structure consisting of a bulk superconducting substrate, a thin insulating layer ($d_I>\xi$), and a superconducting thin film ($d_S\lesssim\lambda$) on top, as illustrated in Fig.~\ref{fig:ML-structure}.

\subsection{Field distributions}
\begin{figure}[h!]
        \centering
        \includegraphics[width=0.5\linewidth]{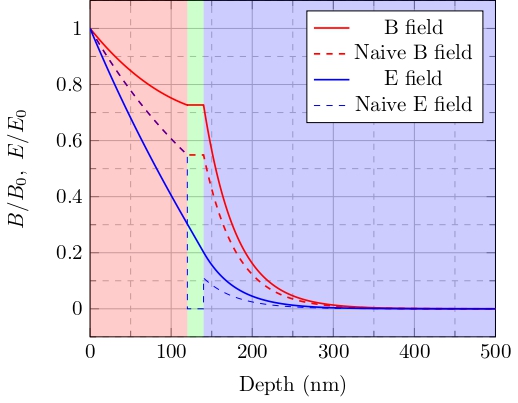}
        \caption{Normalized value of the Electric and Magnetic fields inside the multilayered distribution using the parameters of NbN/I/Nb. The naive fields have been determined by using the exponential decay in the superconductors $e^{-x/\lambda}$ for the magnetic field, and for the electric field $\bm{E}\propto \nabla\times \bm{B}$ due to the Ampère-Maxwell equation. (Color online)}
        \label{fig:NbN-Electric_Magnetic-fields}
    \end{figure}

An external RF electromagnetic field, $\bm{E}$ and $\bm{B}$, is applied parallel to the superconducting surface. 
The field distribution inside the superconductor is determined using the London equations, whereas Maxwell’s equations are applied within the insulating layer.
Following the derivation in Appendix~\ref{sec:App_EM_derivation}, we obtain the magnetic and electric fields inside the multilayer:
\begin{equation}
        B_I=B_0\frac{\cosh\left(\frac{x-d_S}{\lambda_1}\right)-\frac{\lambda_2+d_I}{\lambda_1}\sinh\left(\frac{x-d_S}{\lambda_1}\right)}{\cosh\left(\frac{d_S}{\lambda_1}\right)+\frac{\lambda_2+d_I}{\lambda_1}\sinh\left(\frac{d_S}{\lambda_1}\right)},\quad   E_I=i\omega\lambda_1 B_0 \frac{\sinh\left(\frac{x-d_S}{\lambda_1}\right)-\frac{\lambda_2+d_I}{\lambda_1}\cosh\left(\frac{x-d_s}{\lambda_1}\right)}{\cosh\left(\frac{d_S}{\lambda_1}\right)+\frac{\lambda_2+d_I}{\lambda_1}\sinh\left(\frac{d_S}{\lambda_1}\right)},
    \end{equation}
    \begin{equation}
        B_{II}=\frac{B_0}{\cosh\left(\frac{d_S}{\lambda_1}\right)+\frac{\lambda_2+d_I}{\lambda_1}\sinh\left(\frac{d_S}{\lambda_1}\right)}, \quad E_{II}=\frac{iB_0\omega(x-d_S-d_I-\lambda_2)}{\cosh\left(\frac{d_S}{\lambda_1}\right)+\frac{\lambda_2+d_I}{\lambda_1}\sinh\left(\frac{d_S}{\lambda_1}\right)},
    \end{equation}
    \begin{equation}
        B_{III}=\frac{B_0 \exp\left(-\frac{x-d_S-d_I}{\lambda_2}\right)}{\cosh\left(\frac{d_S}{\lambda_1}\right)+\frac{\lambda_2+d_I}{\lambda_1}\sinh\left(\frac{d_S}{\lambda_1}\right)}, \quad  E_{III}=\frac{-i\omega\lambda_2B_0\exp\left(-\frac{x-d_S-d_I}{\lambda_2}\right)}{\cosh\left(\frac{d_S}{\lambda_1}\right)+\frac{\lambda_2+d_I}{\lambda_1}\sinh\left(\frac{d_S}{\lambda_1}\right)},
    \end{equation}
The magnetic fields reproduce the results from Ref.~\cite{Kubo_2014}, while the electric fields are explicitly shown here for the purposes of this work.
The field distributions are illustrated as solid lines in Fig.~\ref{fig:NbN-Electric_Magnetic-fields} using typical parameters for a NbN/I/Nb multilayer.
For comparison, the dashed lines represent the field distributions obtained from a simple extrapolation of exponential decays in a semi-infinite superconductor, without solving the London and Maxwell equations.
The differences between these two approaches, particularly in the electric field, are critical for accurately evaluating surface resistance and RF losses, as demonstrated in this work for the first time.

\subsection{Vortex penetration field}
The vortex penetration field is determined by both the top and bottom layers, as first calculated in Ref.~\cite{Kubo_2014, Kubo_2017}.
The maximum external magnetic field that the multilayer structure can withstand is primarily limited by the vortex penetration field of the top superconducting layer.
The second superconducting layer is also considered: although the insulating layer protects it from direct vortex entry, the magnetic field reaching this layer must remain below its superheating field to prevent quenching.
Thus, the overall field limit is effectively set by the lower of the two: the vortex penetration field of the top layer, following the derivation in Appendix~\ref{sec:App-Vortex_Calc}, or the superheating field of the bottom layer.
The results can be summarized as follows:
         \begin{equation} \label{eq:Max-field}
        B_v=\left\lbrace\begin{aligned}
            &\frac{\phi_0}{4\pi \lambda_1 \xi_1}\frac{\cosh\left(\frac{d_S}{\lambda_1}\right)+\frac{\lambda_2+d_I}{\lambda_1}\sinh\left(\frac{d_S}{\lambda_1}\right)}{\sinh\left(\frac{d_S}{\lambda_1}\right)+\frac{\lambda_2+d_I}{\lambda_1}\cosh\left(\frac{d_S}{\lambda_1}\right)}\quad \text{if} \ B_{sh}^{(\text{Bulk},S_2)}>\frac{\phi_0}{4\pi \lambda_1 \xi_1}\frac{1}{\sinh\left(\frac{d_S}{\lambda_1}\right)+\frac{\lambda_2+d_I}{\lambda_1}\cosh\left(\frac{d_S}{\lambda_1}\right)},\\
            &\left[\cosh\left(\frac{d_S}{\lambda_1}\right)+\frac{\lambda_2+d_I}{\lambda_1}\sinh\left(\frac{d_S}{\lambda_1}\right)\right]B_{sh}^{(\text{Bulk},S_2)} \quad \text{else,}
        \end{aligned}\right.
    \end{equation}
    %Which leads to the same result as Ref.~\cite{Kubo_2014}.
where $\phi_0=2.07\cdot 10^{-15}$~Wb is the magnetic flux quantum, which has a topological origin.
    
\subsection{Surface resistance and RF loss}
The power loss per area under an external RF field $\bm{H}$\footnote{In Section~\ref{sec:Num_Results}, we will select the highest possible magnetic field that can be applied to the multilayer to find the thermal instability at the highest possible quench field.} and $\bm{E}$ can be given by
    \begin{equation} \label{eq:power-loss}
        P_s=\frac{1}{2}R_s|\bm{H}|^2=\frac{\sigma_1}{2}\int_0^\infty {\rm{d}}x \ |\bm{E}(x)|^2,
    \end{equation}
Equation~\eqref{eq:power-loss} leads to the surface resistance of the multilayer being~\cite{Gurevich_2017}
   \begin{equation}\label{eq:Surface-Resistance}
       R_s=\frac{\mu_0^2}{B_0^2}\left\lbrace \sigma_{1;S_1}\int_0^{d_S} {\rm d}x \ |E_I(x)|^2+2q+\sigma_{1;S_2}\int_{d_S+d_I}^\infty {\rm d}x \ |E_{III}(x)|^2\right \rbrace.
   \end{equation}
Here, $q$ represents the power loss due to dielectric dissipation, given by
   \begin{equation}
       q=\omega \varepsilon \tan\delta \int_{d_S}^{d_I+d_S}{\rm d}x \ |E_{II}(x)|^2,\qquad \tan\delta=\frac{\omega \text{Im}(\varepsilon)+\sigma_I}{\omega\text{Re}(\varepsilon)},
   \end{equation}
where $\tan\delta$ is the loss tangent of the insulating layer, and $\varepsilon$ and $\sigma_{I}$ are its complex electric permittivity and conductivity, respectively

After straightforward calculations, Eq.~\eqref{eq:Surface-Resistance} can be expressed as
   \begin{equation}
     \begin{aligned}
       R_{s} & \equiv R_{s;S_1} + R_{s;I} + R_{s;S_2} \\
                  & \equiv \tilde{R}_{s;S_1}D_{1} + R_{s;I} + \tilde{R}_{s;S_2}D_{2},
       \end{aligned}
   \end{equation}
where $\tilde{R}_{s, S_k}$ denotes the bulk surface resistance of layer $S_k$, and $D_{k}$ ($k=1,2$) is an \textit{attenuation factor} arising from the layered structure.
The first attenuation factor, $D_{1}$, originates from the finite thickness of the top superconducting layer, whereas the second factor, $D_{2}$, accounts for the screening of the fields by the first layer before reaching the second superconducting layer:
     \begin{equation} \label{eq:layer 1}
    \begin{aligned}
        D_1(d_I,d_S)&=\frac{1}{2\lambda_1^3\left[\cosh\left(\frac{d_S}{\lambda_1}\right)+\frac{\lambda_2+d_I}{\lambda_1}\sinh\left(\frac{d_S}{\lambda_1}\right)\right]^2}\left\lbrace 2d_S[(\lambda_2+d_I)^2-\lambda_1^2]-2\lambda_1^2(\lambda_2+d_I)\right.\\
        &\left.+2\lambda_1^2(\lambda_2+d_I)\cosh\left(\frac{2d_S}{\lambda_1}\right)
         +\lambda_1[\lambda_1^2+(\lambda_2+d_I)^2]\sinh\left(\frac{2d_S}{\lambda_1}\right)\right\rbrace,
    \end{aligned}
    \end{equation}
    \begin{equation} \label{eq:layer 3}
        D_2(d_I,d_S)=\frac{1}{\left[\cosh\left(\frac{d_S}{\lambda_1}\right)+\frac{\lambda_2+d_I}{\lambda_1}\sinh\left(\frac{d_S}{\lambda_1}\right)\right]^2}.
    \end{equation}
Finally, the effective resistance from the insulating layer is denoted by
    \begin{equation}
       R_{s;I} \equiv2\frac{\mu_0^2}{B_0^2}q=\frac{2\mu_0^2\omega^3\varepsilon_0\varepsilon_r\tan(\delta)[(d_I+\lambda_2)^3-\lambda_2^3]}{3\left[\cosh\left(\frac{d_S}{\lambda_1}\right)+\frac{\lambda_2+d_I}{\lambda_1}\sinh\left(\frac{d_S}{\lambda_1}\right)\right]^2}.
    \end{equation}

One unknown parameter in the equations is the real part of the optical conductivity, $\sigma_1$. 
For simple $s$-wave superconductors, Eq.~\eqref{eq:BCS-MB-simple} can be rewritten through the Mattis-Bardeen theory in the local limit~\cite{Gurevich_2017}
    \begin{equation}
        \sigma_1=\frac{4\sigma_0\Delta(T)}{2k_B T}\ln\left(\frac{4e^{-\gamma_E}k_BT}{\hbar \omega}\right)e^{-\Delta(T)/k_BT}. \label{eq:BCS-MB_local_limit}
    \end{equation}
Reflecting the complex gap structure, the optical conductivity of IBS was evaluated via a numerical integral introduced by one of the authors in Ref.~\cite{10601306}, based on a phenomenological model of an $s^{\pm}$-wave superconductor proposed by Nagai~\cite{Nagai_2008}.
Owing to their different physical nature and potential applications, $d$-wave superconductors are not considered in this work.   

\section{Results} \label{sec:Num_Results}
We calculated the maximum field and surface resistance as functions of the insulator thickness, $d_I$, and the top layer thickness, $d_S$, for multilayer structures based on conventional superconductors and IBS.
The assumed material parameters are summarized in Appendix~\ref{sec:Appendix_Num_results}.
The environmental conditions are set to $T = 2$~K and $\omega/2\pi = 1.3$~GHz, which are the most commonly used in particle accelerator applications.
Nb is assumed as the bottom layer for most examples at $T = 2$~K; however, alternative substrates are also considered for potential improvements and operation at higher temperatures.

\subsection{NbN/I/Nb Multilayer Structure}
\begin{figure}[h!]
 \centering
    \begin{subfigure}[t]{0.5\textwidth}
        \centering
        \includegraphics[width=0.9\linewidth]{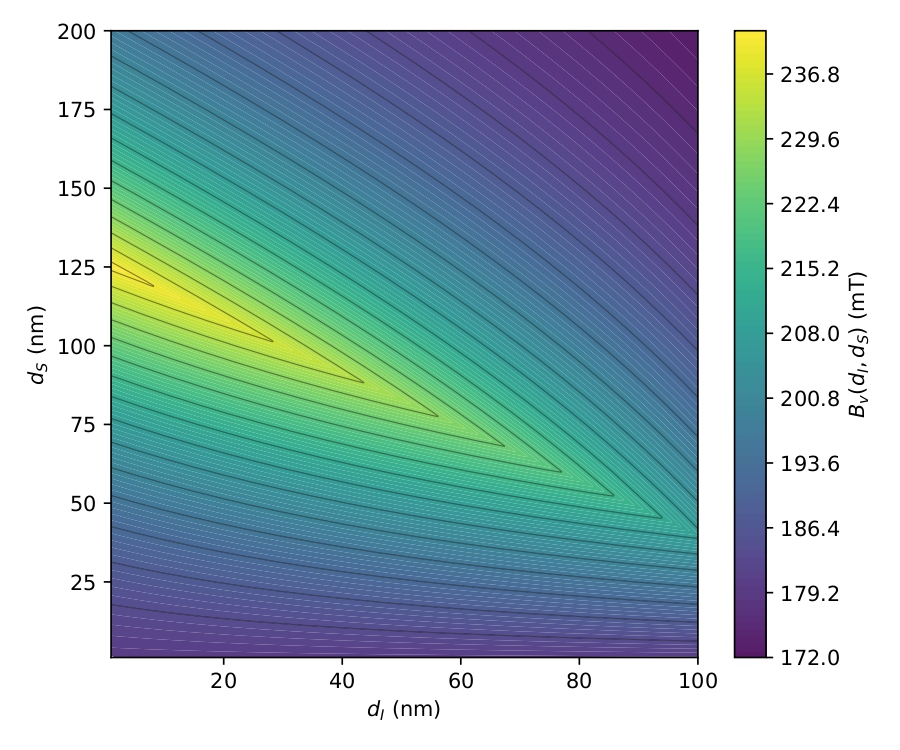}
        \caption{Maximum magnetic field applicable to a NbN/I/Nb multilayer structure, when we plot it against the thickness of the superconductor $d_S$ and the thickness of the insulator $d_I$.}
        \label{fig:Bv_NbN_fig}
    \end{subfigure}%
    ~ 
    \begin{subfigure}[t]{0.5\textwidth}
        \centering
        \includegraphics[width=0.9\linewidth]{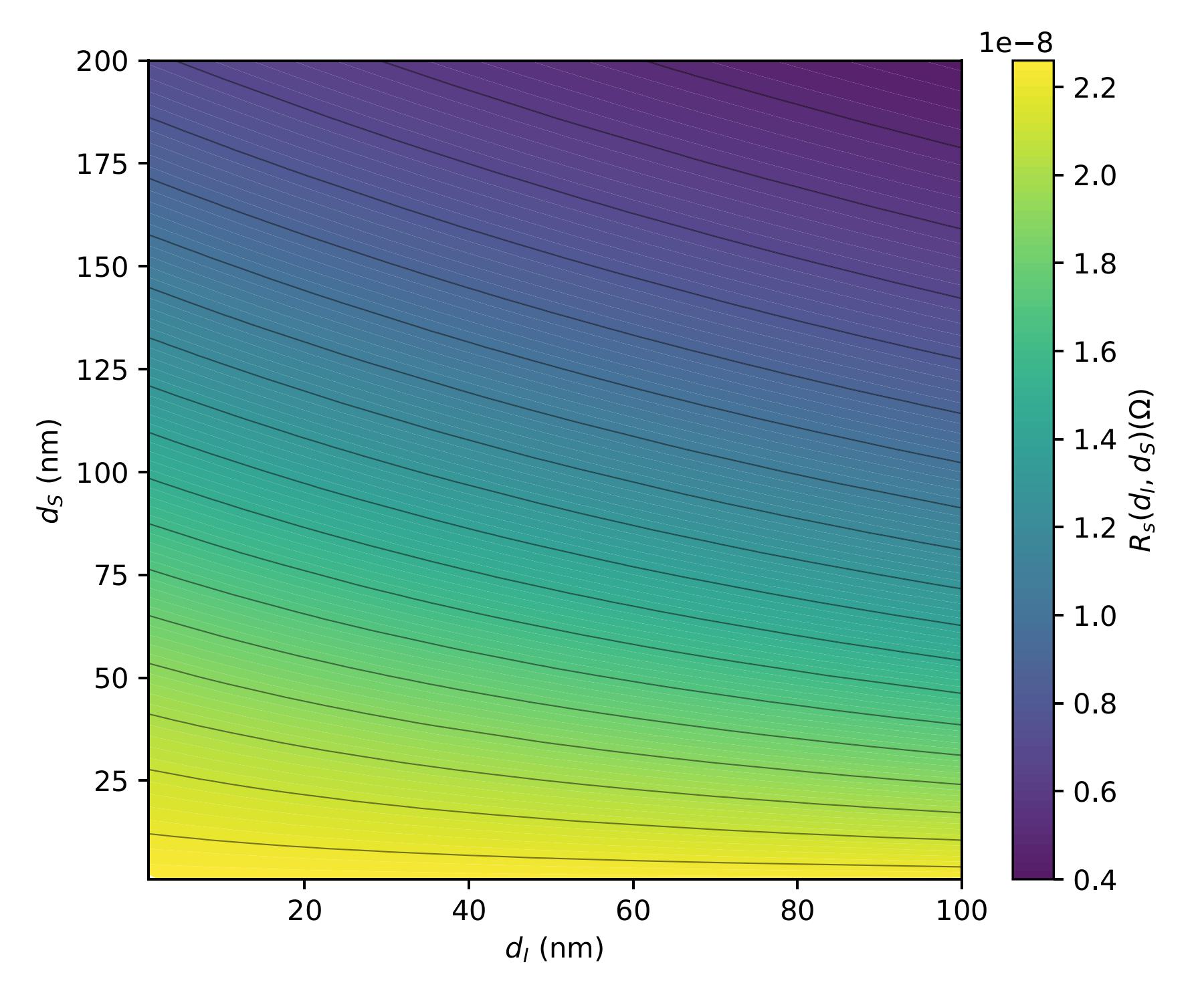}
        \caption{Surface Resistance dependence of the multilayer on the thickness of the first SC layer $d_S$, and the insulator layer $d_I$.}
        \label{fig:Rs_NbN_fig}
    \end{subfigure}
    \caption{Results for the NbN/I/Nb multilayer structure. Figure (a): maximum applicable field of the multilayer. Figure (b): Surface resistance of the multilayer. (Color online)}
    \label{fig:NbN-Nb_results}
    \end{figure}

The maximum field of NbN/I/Nb multilayers has been extensively studied both theoretically~\cite{Kubo_2014} and experimentally~\cite{katayama:KEK-NbN,HZB:2022jjd}.
In this study, we extend these investigations by including the surface resistance, as shown in Fig.~\ref{fig:NbN-Nb_results}, and the corresponding power loss.
    
Figure~\ref{fig:Bv_NbN_fig} reproduces the results of Kubo~\cite{Kubo_2014}.
As observed in Fig.~\ref{fig:Bv_NbN_fig}, the maximum field is achieved when the insulating layer is absent ($d_I \to 0$), as suggested by Ref.~\cite{Asaduzzaman_2024}.
However, achieving $d_I \to 0$ is practically challenging due to two effects: defects and Josephson vortices.
Defects in a realistic top layer locally weaken the Bean-Livingston barrier~\cite{Bean:1964zz}, which can be mitigated by including a thin insulating layer ($d_I > 0$) that protects against RF vortices penetrating into the substrate.
Additionally, if $\xi_0 > d_I$, the Josephson effect between the two superconducting layers can trap Josephson vortices, causing additional RF losses.

Excluding the $d_I = 0$ case, Fig.~\ref{fig:Bv_NbN_fig} shows that the maximum field is $B_v = 243.3$~mT when the NbN layer thickness is $d_S = 125$~nm and the insulating layer thickness is $d_I = 5$~nm.
For the same parameters, Fig.~\ref{fig:Rs_NbN_fig} indicates a reasonably low surface resistance of\footnote{To denote the surface resistance of a semi-infinite bulk material, we use a tilde, $\tilde{R}s$, as in the theory section.}
$R_s = 0.55 \cdot \tilde{R}_{s,\text{Nb}} = 12.40~\text{n}\Omega$.
We also considered the contribution of a typical insulator, Al$_2$O$_3$, which, with the optimal thickness parameters, contributes $R_{s,I} = 4.0 \times 10^{-22}~\Omega$, negligibly small compared to the surface resistance of the superconducting layers.
Finally, the optimal parameters yield a power loss per unit area of $P_{\text{NbN}/\text{Nb}} = 232.4~\text{W/m}^2$. 

\subsection{Nb$_3$Sn/I/Nb Multilayer structure}
\begin{figure}[h!]
 \centering
    \begin{subfigure}[t]{0.5\textwidth}
        \centering
        \includegraphics[width=0.9\linewidth]{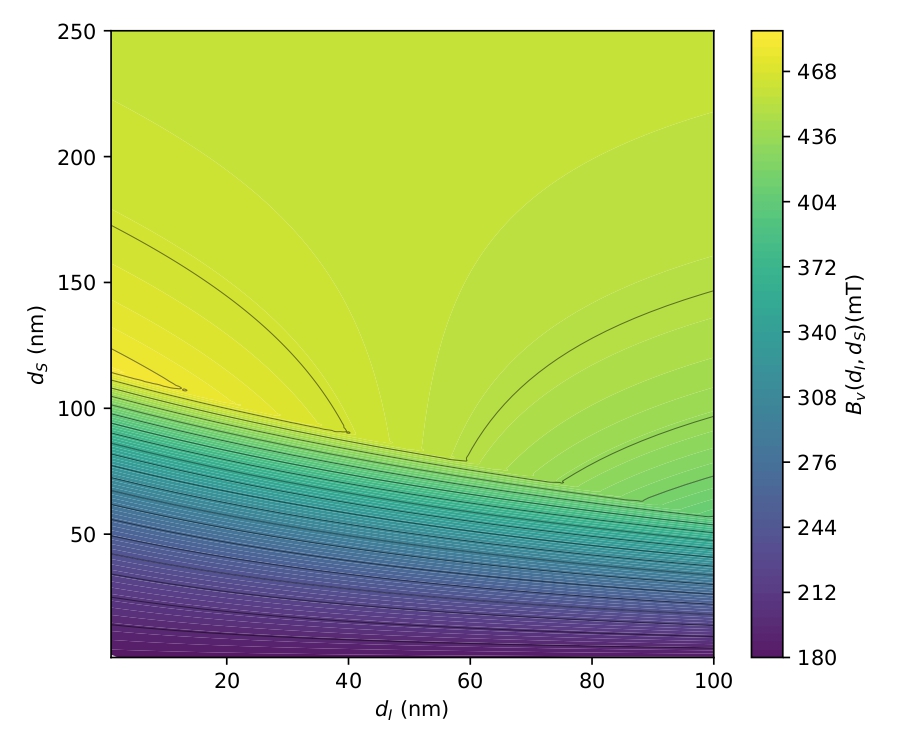}
        \caption{Maximum magnetic field applicable to a Nb$_3$Sn/I/Nb multilayer structure, when we plot it against the thickness of the superconductor $d_S$ and the thickness of the insulator $d_I$.}
        \label{fig:Bv_Nb3Sn_fig}
    \end{subfigure}%
    ~ 
    \begin{subfigure}[t]{0.5\textwidth}
        \centering
        \includegraphics[width=0.9\linewidth]{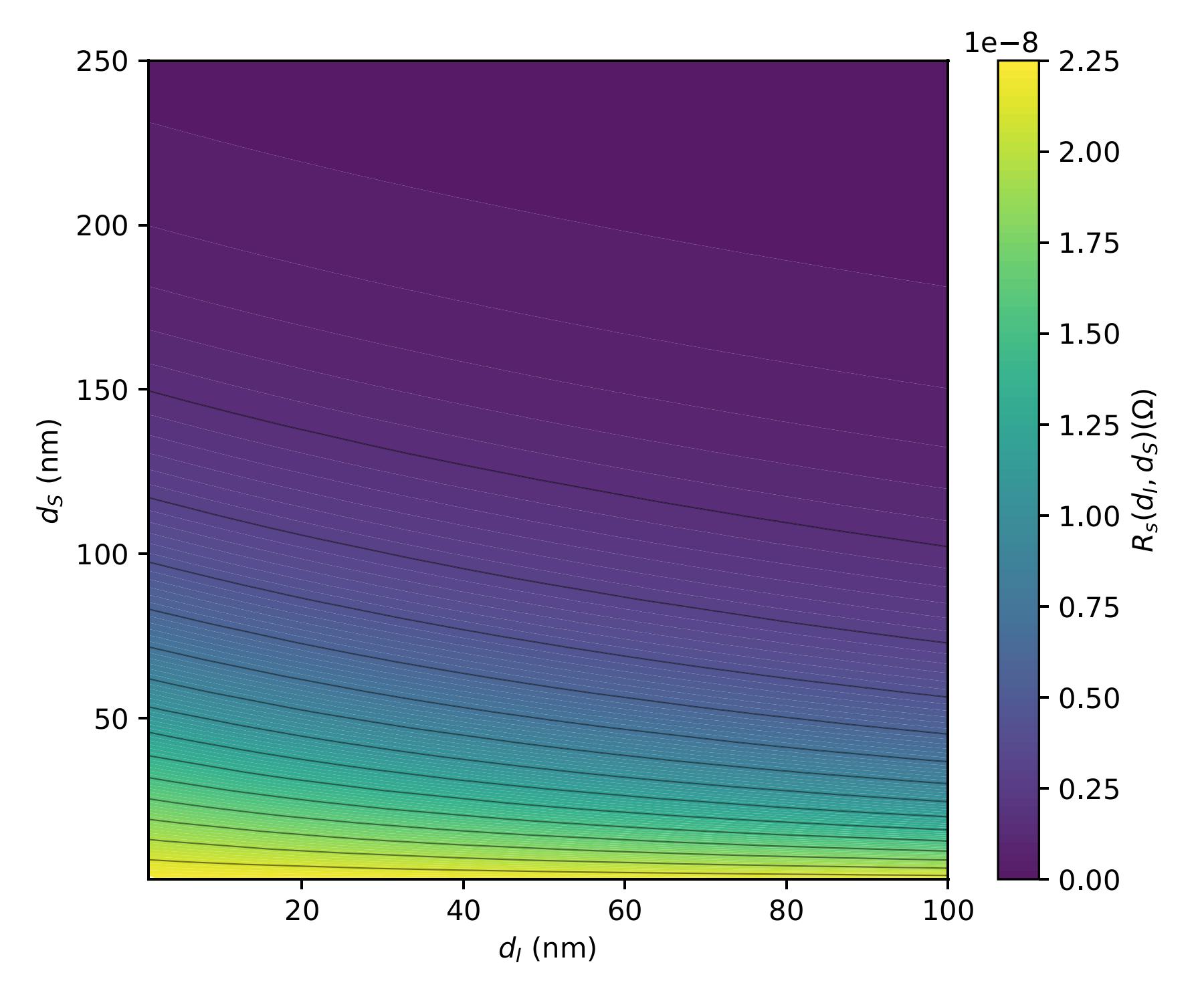}
        \caption{Surface Resistance dependence of the multilayer on the thickness of the first SC layer $d_S$, and the insulator layer $d_I$.}
        \label{fig:Rs_Nb3Sn_Nb}
    \end{subfigure}
    \caption{Results for the Nb$_3$Sn/I/Nb multilayer structure. Figure (a): maximum applicable field of the multilayer. Figure (b): Surface resistance of the multilayer. (Color online)}
    \label{fig:Nb3Sn-Nb_results}
    \end{figure}

The same numerical analysis provides the maximum field and surface resistance for Nb$_3$Sn/I/Nb, the most commonly studied candidate for a promising multilayer structure, as shown in Fig.~\ref{fig:Nb3Sn-Nb_results}.
This system has been investigated both within the framework of Ginzburg-Landau theory~\cite{10.3389/femat.2023.1246016} and through experiments~\cite{Ito:2019Nb3Sn}.

In this case, the maximum field is achieved when the superconducting layer thickness is $d_S = 110$~nm and the insulating layer thickness is $d_I = 10$~nm.
This corresponds to a maximum field of $B_v = 480.8$~mT and a multilayer surface resistance of $R_s = 0.14 \cdot \tilde{R}_{s,\text{Nb}} = 3.09~\text{n}\Omega$.
These values result in a power loss per unit area of $P_{\text{Nb}_3\text{Sn}/\text{Nb}} = 226.3~\text{W/m}^2$.
As expected, this represents a significant improvement over the NbN/I/Nb structure in terms of both $R_s$ and $B_v$.

\subsection{FeSe/I/Nb Multilayer Structure}
\begin{figure}[h!]
 \centering
    \begin{subfigure}[t]{0.5\textwidth}
        \centering
        \includegraphics[width=0.9\linewidth]{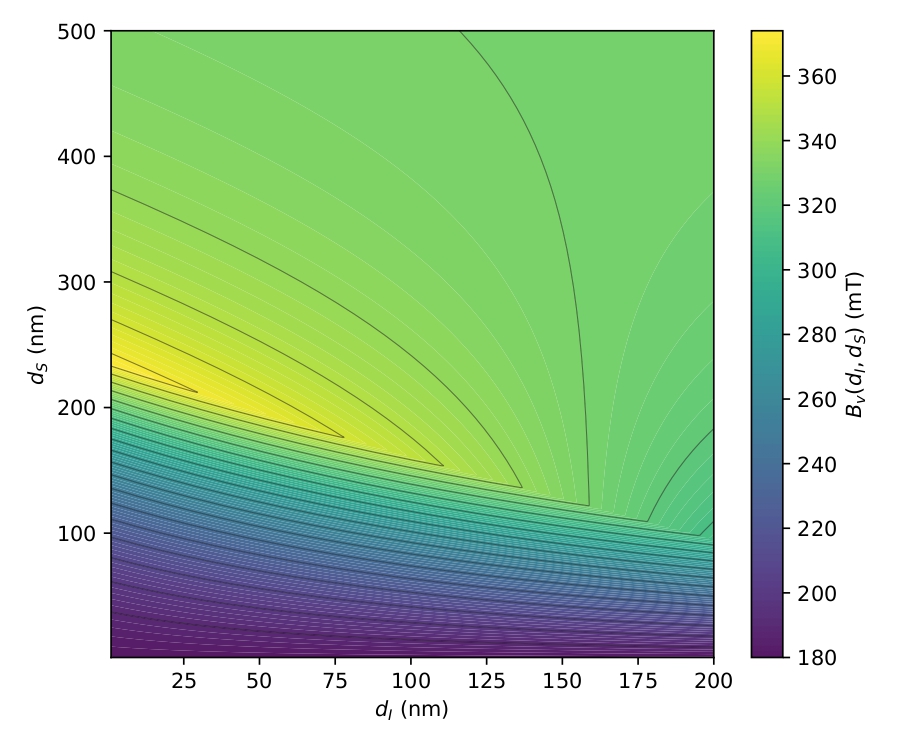}
        \caption{Maximum magnetic field applicable to a FeSe/I/Nb multilayer structure, when we plot it against the thickness of the superconductor $d_S$ and the thickness of the insulator $d_I$.}
        \label{fig:Bv_FeSe}
    \end{subfigure}%
    ~ 
    \begin{subfigure}[t]{0.5\textwidth}
        \centering
        \includegraphics[width=0.9\linewidth]{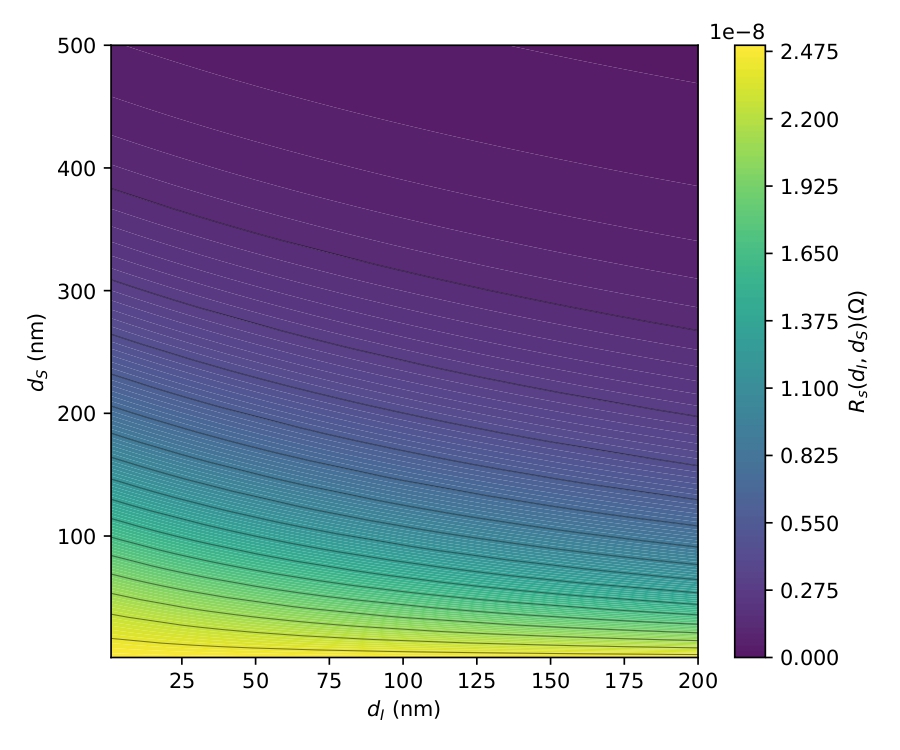}
        \caption{Surface Resistance dependence of the multilayer on the thickness of the first SC layer $d_S$, and the insulator layer $d_I$.}
        \label{fig:Rs_FeSe_Nb}
    \end{subfigure}
    \caption{Results for the FeSe/I/Nb multilayer structure. Figure (a): maximum applicable field of the multilayer. Figure (b): Surface resistance of the multilayer. (Color online)}
    \label{fig:FeSe-Nb_results}
    \end{figure}

Previous work by one of the authors~\cite{10601306} showed that the surface resistance of bulk pnictides is not promising due to their large penetration depth, despite their potential for high-field and/or high-temperature applications.
In this study, a simpler IBS, FeSe, is considered, as FeSe films have already been experimentally realized.
The numerical analysis yields the maximum field and surface resistance, as shown in Fig.~\ref{fig:FeSe-Nb_results}.
    
The optimal parameters, $d_S = 215.15$~nm and $d_I = 25$~nm, yield a maximum field of $B_v = 370$~mT.
The corresponding surface resistance for these layer thicknesses is $R_s = 0.24 \cdot \tilde{R}_{s,\text{Nb}} = 5.354~\text{n}\Omega$.
Finally, the power loss per unit area for this configuration is $P_{\text{FeSe/Nb}} = 232.1~\text{W/m}^2$. 
    
\subsection{FeSe/I/Nb$_3$Sn}
\begin{figure}[h!]
 \centering
    \begin{subfigure}[t]{0.5\textwidth}
        \centering
        \includegraphics[width=0.9\linewidth]{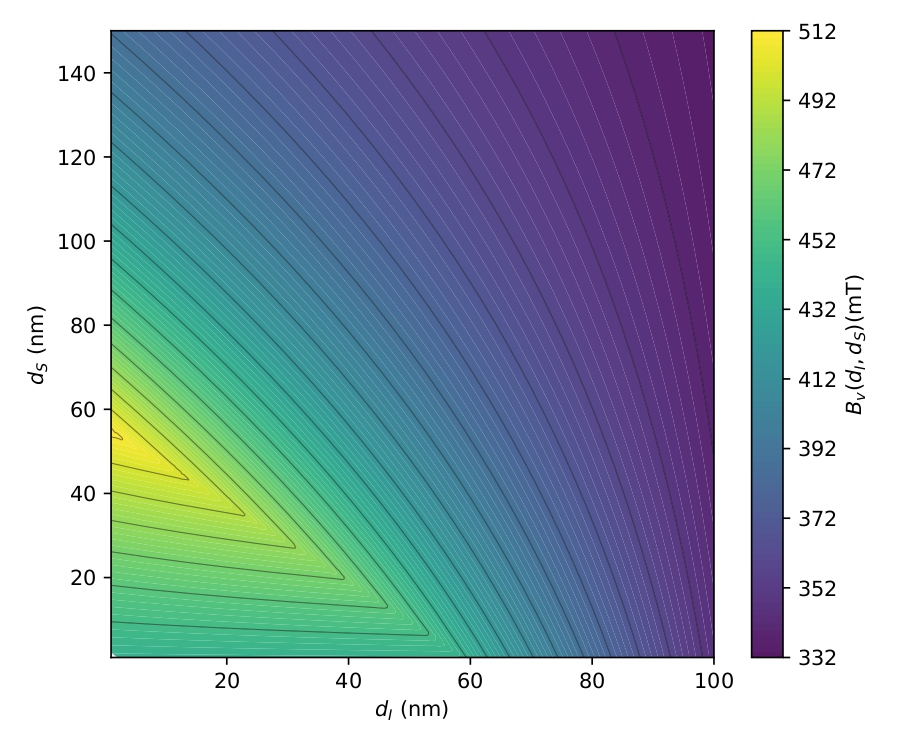}
        \caption{Maximum magnetic field applicable to a FeSe/I/Nb$_3$Sn multilayer distribution, when we plot it against the thickness of the superconductor $d_S$ and the thickness of the insulator $d_I$. (Color online)}
        \label{fig:Bv_FeSe_Nb3Sn_fig}
    \end{subfigure}%
    ~ 
    \begin{subfigure}[t]{0.5\textwidth}
        \centering
        \includegraphics[width=0.9\linewidth]{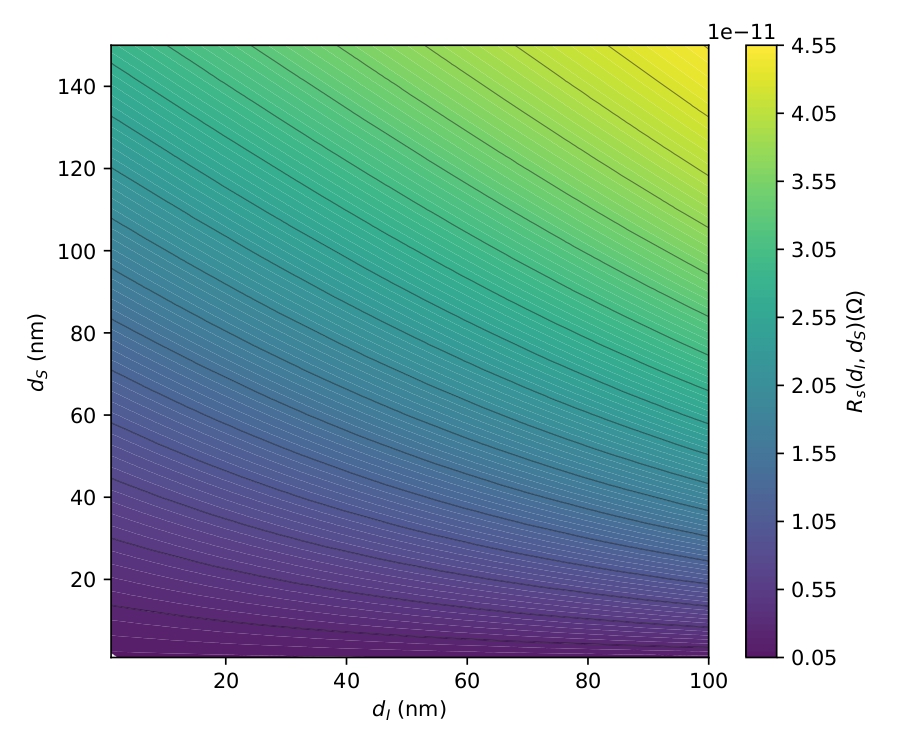}
        \caption{Surface Resistance dependence of the multilayer on the thickness of the first SC layer $d_S$, and the insulator layer $d_I$.}
        \label{fig:Rs-FeSe-Nb3Sn}
    \end{subfigure}
    \caption{Results for the FeSe/I/Nb$_3$Sn multilayer structure. Figure (a): maximum applicable field of the multilayer. Figure (b): Surface resistance of the multilayer.}
    \label{fig:FeSe-Nb3Sn_results}
    \end{figure}

To explore both substantially enhanced performance and the potential for higher-temperature operation, FeSe was considered as a top layer deposited on Nb$_3$Sn, serving as an extreme example.
This configuration can be realized on the established Nb$_3$Sn/Nb bilayer via Sn vapor deposition, since the bulk Nb$_3$Sn layer is sufficiently thicker than $\lambda$.
The numerical analysis provides the maximum field and surface resistance, as shown in Fig.~\ref{fig:FeSe-Nb3Sn_results}.
  
As shown in Fig.~\ref{fig:FeSe-Nb3Sn_results}, the maximum field that can be applied is $B_v = 508.3$~mT when the superconducting and insulating layers have thicknesses of $d_S = 51$~nm and $d_I = 5$~nm, respectively.
This configuration results in a multilayer surface resistance of $R_s = 4.18 \times 10^{-4} \cdot \tilde{R}_{s,\text{Nb}} = 9.31 \times 10^{-3}~\text{n}\Omega$, corresponding to a power loss per unit area of $P_{\text{FeSe}/\text{Nb}_3\text{Sn}} = 0.75~\text{W/m}^2$. 

\section{Discussions} 
\subsection{Influence of the substrate}
\begin{figure}[h!]
 \centering
    \begin{subfigure}[t]{0.5\textwidth}
        \centering
        \includegraphics[width=0.9\linewidth]{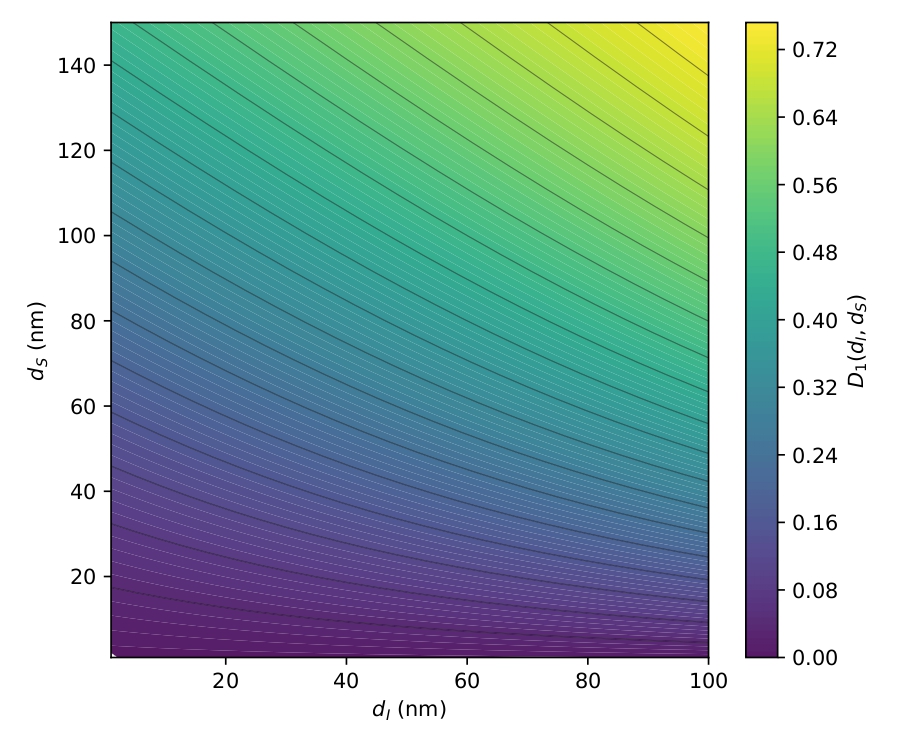}
        \caption{Attenuation factor $D_1(d_I,d_S)$ of the surface resistance in the first superconducting layer for a FeSe/I/Nb$_3$Sn multilayer. (Color online)}
    \end{subfigure}%
    ~ 
    \begin{subfigure}[t]{0.5\textwidth}
        \centering
        \includegraphics[width=0.9\linewidth]{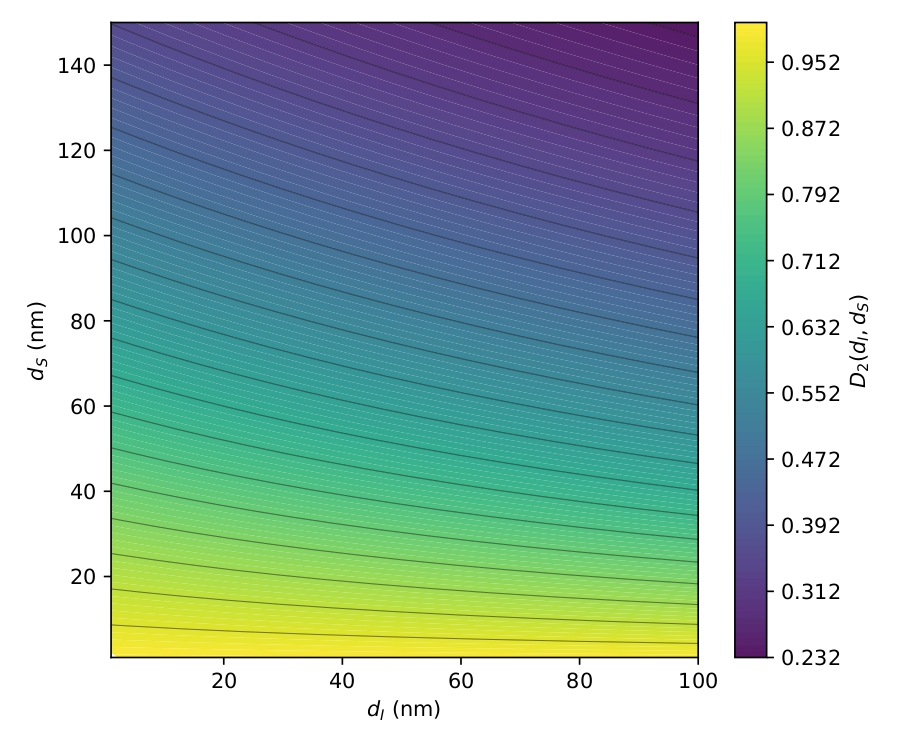}
        \caption{Attenuation factor $D_2(d_I,d_S)$ of the surface resistance in the second superconducting layer for a FeSe/I/Nb$_3$Sn multilayer.}
        \label{fig:D2-FeSe-Nb3Sn}
    \end{subfigure}
    \caption{Attenuation factors $D_1(d_I,d_S)$ and $D_2(d_I,d_S)$ in the FeSe/I/Nb$_3$Sn multilayer.}
    \label{fig:Attenuation-factors}
    \end{figure}

As shown in Fig.~\ref{fig:Rs-FeSe-Nb3Sn}, the surface resistance of this multilayer behaves differently from that of the other proposed materials (Figs.~\ref{fig:Rs_NbN_fig},~\ref{fig:Rs_Nb3Sn_Nb},~\ref{fig:Rs_FeSe_Nb}).
This is due to the bulk surface resistance of FeSe being larger than that of Nb$_3$Sn, as can be inferred from the attenuation factors in Eqs.~\eqref{eq:layer 1} and \eqref{eq:layer 3}, plotted in Fig.~\ref{fig:Attenuation-factors}.
For typical multilayer parameters, the surface resistance is generally dominated by the substrate rather than the top layer.
This behavior has not been reported previously, as multilayers are usually designed to maximize the critical field of the top layer without considering surface resistance.
Consequently, depending on the properties of individual materials and the layer configuration, achieving the highest quench field may be accompanied by increased RF losses.
The simultaneous consideration of vortex penetration fields and surface resistance, as proposed in this work, introduces a new optimization strategy for future multilayers employing various non-conventional superconductors.

\subsection{Comparison of various multilayers}
We compare the optimal layer parameters, along with the corresponding $B_v$ and $R_s$, for these multilayers and bulk Nb in Table~\ref{tab:my_label}.
\begin{table}[h!]
    \centering
    \begin{tabular}{|c|c|c|c|c|c|}\hline
         & FeSe/I/Nb & NbN/I/Nb & Nb$_3$Sn/I/Nb & FeSe/I/Nb$_3$Sn & Nb \\\hline
         $B_v$(mT)&370&243.3&480.8&508.3&180\\\hline
         %$R_s(\Omega)$&$5.354\cdot 10^{-9}$&$1.240\cdot 10^{-8}$&$3.094\cdot 10^{-9}$&$2.770\cdot 10^{-12}$&$2.226\cdot10^{-8}$\\\hline
         $R_s(n\Omega)$&$5.354$&$12.40$&$3.094$&$2.770\cdot 10^{-3}$&$22.26$\\\hline
         $P(B_v)(\text{W}\text{m}^2)$&232.1&232.4&226.3&0.75&232.1\\\hline
          $d_S$(nm)&215.15&125&110&51&0\\\hline
         $d_I$(nm)&25&5&10&5&0\\\hline
    \end{tabular}
    \caption{Comparison between the optimum multilayer structures made out of different materials and bulk Nb.}
    \label{tab:my_label}
\end{table}

Table~\ref{tab:my_label} confirms that Nb$_3$Sn/I/Nb provides the best performance on an Nb substrate.
However, despite the recent successful demonstration of Nb$_3$Sn/Nb cavities in a cryomodule~\cite{Eremeev_2025,Posen_2021}, Nb$_3$Sn is mechanically brittle, limiting its tuning capabilities.
The predicted RF performance of FeSe/I/Nb is comparable to that of Nb$_3$Sn/I/Nb, and its metallic nature could potentially allow mechanical tuning to synchronize cavities for particle accelerator applications. 

This paper has focused on FeSe/I/Nb due to its technicial readiness, as demonstrated in small-scale samples~\cite{FeSe:CAS}.
Other pnictide materials, such as SrFeAs, have been developed for superconducting wires but not for SRF applications.
The key technological challenge for multilayer deposition is the safe handling of arsenide.
Since arsenide compounds are routinely used in the semiconductor industry, e.g., in GaAs, we argue that there should be no fundamental obstacles to depositing pnictide layers once the appropriate infrastructure is established.
Based on the Nagai model describing the $s^\pm$-wave superconducting properties of pnictides, these materials not only exhibit higher critical temperatures, $T_c$, but also feature an optical conductivity, $\sigma_1$, that increases more slowly with temperature~\cite{10601306}, making them potentially useful at temperatures equal to or above 4~K.
Nevertheless, at 4~K, Nb$_3$Sn still outperforms pnictides. 
    
Moreover, as discussed in the previous section, most of the surface resistance arises from the second superconducting layer.
Therefore, to reduce the surface resistance, one can either accept a partial reduction in the maximum field, $B_v$, by adjusting the layer thicknesses, $d_I$ and $d_S$, or replace the second Nb layer with a material having superior properties, such as NbN or Nb$_3$Sn, which exhibit lower surface resistance.

Despite the widely different values of $B_v$ and $R_s$, it is noteworthy that the power flux remains roughly the same across all materials, except for FeSe/I/Nb$_3$Sn.
This suggests that optimizing thermal resistance, particularly the Kapitza resistance at the interfaces of multilayers, may play a crucial role in the overall performance.
This aspect is briefly addressed in Sec.~\ref{sec:imperfect}.

\subsection{Comparison with experimental data}
There exists limited experimental data on the multilayer structure in the literature.
They can be categorized into two:
\begin{description}
\item [DC measurement] R. Katayama et al.~\cite{katayama:srf2019-thfua2} demonstrated the enhancement of the first vortex penetration field in a small sample of NbN/SiO$_{2}$/Nb by using the 3rd harmonic generation. They saw good agreement between their data and the multilayer model. Importantly, they employed magnetic fields oscillating at 15~kHz, and thus it was effectively a DC measurement. Md Asaduzzaman et al.~\cite{Asaduzzaman_2024} visualized the DC magnetic field profile inside the layered structure by using the $\mu$SR technique. They observed the suppression of DC surface current as predicted by the multilayer model. For the iron-based superconductors, Z. Lin et al.~\cite{Lin_2021} observed enhancement of the lower and upper DC critical fields of FeSe-coated Nb. From these experimental evidences, one can assume that the enhancement of the vortex penetration field by the multilayer structure is more or less established at least in the case of DC. The remaining challenge is its response to RF.
\item [RF measurement] This is a more crucial comparison to the simultaneous optimization of vortex penetration field and RF surface resistance, firstly proposed by this work. D. Tkhonov et al.~\cite{tikhonov:srf2021-supfdv006} presented the surface resistance of various multilayer structures, such as NbN/AlN/Nb, evaluated with a flat sample measurement by a quadrupole resonator (QPR). However, the results are dominated by large residual resistance, as is usually the case in QPR measurement. This might be some common systematic effect of the QPR measurement itself. Since our theory predicts ideal surface resistance, this data is not immediately available for quantitative comparisons. The large residual resistance could be partly associated with the potential thermal issues of the multilayer structure as discussed in~\ref{sec:imperfect}.
\end{description}
From the above argument, we conclude that explicit comparison to the existing data may not be very relevant at this stage. One needs to collect more RF measurement data in the near future.

\subsection{Effect of imperfections} \label{sec:imperfect}
Some imperfections in the thin films and the interface between layers can be a practical issue in the multilayer structure. 
%The complete modeling of any imperfections is outside the scope of this manuscript.
For example, inter-diffusion is a challenge during the layer deposition and needs to be technically suppressed by a passivation layer.
The proximity effect cannot be addressed by the current formalism with the London equations.
This may be studied in future work with the Eilenberger equation. 
For the theoretical prediction on imperfections, a recent work by A. Gobeyn et al.~\cite{gobeyn2026numericalqualityfactorstatistics} addresses inhomogeneous layer thickness over the cavity surface by using Gaussian random fields.
They showed that such an effect has a marginal impact on the electromagnetic properties of the multilayer-coated cavities.
The very first multilayer cavity is being prepared in DESY~\cite{ML-cavity}.

The complete calculation of the thermal effect through the imperfect interface is outside the scope of this work.
Instead, this paper provides the power loss per unit area as an input parameter for future thermal studies.
The contact thermal resistance due to Kapitza resistance is a long-lasting concern in the thin-film SRF community. 
This may partially explain the high residual resistance in the RF measurement on QPR samples, as QPR is based on a calorimetric method that employs DC heaters and temperature sensors behind the sample to estimate the RF loss.
Although it is hard to precisely model the microscopic features of the interface, macroscopic contact resistance can be phenomenologically modeled in the COMSOL Multiphysics~\cite{COMSOL} simulation. 
Obtaining reliable contact thermal resistance data from a sample measurement~\cite{WENSKAT20251354694} may be the next research direction for multilayer structures. 
Another approach to handle the thermal contact issue in the interface was theoretically proposed by V. Palmieri and R. Vaglio~\cite{palmieri16} and experimentally studied by one of the authors~\cite{calatroni16} for niobium-coated copper cavities.
This model could phenomenologically explain the nonlinear increase in surface resistance due to imperfect contact resistance between two layers.
The Palmieri model converts the probability density distribution of imperfect thermal resistance into surface resistance as a function of RF fields.
Lacking sufficient data prevents us from applying this model to the multilayer structure.

Unlike thermal contact resistance, electrical contact is usually subdominant in the SRF application because the RF current is parallel to the interface. The Josephson effect in a bilayer is weakly destroyed at low enough temperatures, even when the contact is poor,
from a theoretical calculation~\cite{PhysRevB.73.134504}.
Therefore, as long as the temperature is low enough, the contact electrical resistance is small compared to the BCS surface resistance.

\section{Conclusion} 
In this paper, we have shown how multilayers can enhance the quench field compared to bulk materials such as Nb.
Using this framework, based on a classical description of superconductivity, we systematically presented a method to calculate the maximum field applicable to a multilayer structure, its surface resistance, and the power loss per unit area.
We also discussed the potential of IBS in the development of SRF cavities operating at temperatures above 4~K, or even at lower temperatures, using FeSe as an illustrative example.
Extending these results to more precise measurements, including AsFeSe, represents a promising direction for future work.    

%%%%%%%%%%%%%%%%%%%%%%%%%%%%%%%%%%%%%%%%%%%%%%%%%%%%%%%%%%%%%%%%%%%%%%%%%%%%%%%%%%
\ack{}
We would like to thank A. Perez Ruiz, D. Longuerverne, C. Boutelaa, C. Cerna, O. Quaranta, T. Guruswamy, H. Hu, and D. Bafia for useful discussions.
We are also grateful towards ANL and FNAL hospitality, as part of this work was carried out at these institutions. 
This work was supported by the CNRS-UChicago IRC grant, FACCTS funding at UChicago, and the European Union’s Horizon Europe research and innovation programme under Grant Agreement No. 101086276 (EAJADE).

%%%%%%%%%%%%%%%%%%%%%%%%%%%%%%%%%%%%%%%%%%%%%%%%%%%%%%%%%%%%%%%%%%%%%%%%%%%%%%%%%%

   \appendix
   \section{Derivation of the EM field inside a multilayer} \label{sec:App_EM_derivation}
   Assuming that the external AC field is an EM wave of $\omega$ frequency
   \begin{equation}
       \bm{E}=(0,E,0)e^{-i\omega t },\qquad \bm{B}=(0,0,B)e^{-i\omega t}
   \end{equation}
   The Maxwell and London equations that describe the magnetic field inside the multilayer are:
   \begin{equation}
   \left\lbrace
    \begin{aligned}
        \frac{d^2 B}{dx^2}&=\frac{B}{\lambda_1^2} \qquad \text{if} \ x\in[0,d_S),\\
        \frac{d^2 B}{dx^2}&=-\frac{\omega^2 \varepsilon_r}{c^2}B\qquad \text{if} \ x\in [d_S,d_S+d_I),\\
        \frac{d^2 B}{dx^2}&=\frac{B}{\lambda_2^2} \qquad \text{if} \ x\in [d_S+d_I,\infty).
    \end{aligned}\right.
   \end{equation}
   And the electric field can be determined using the Ampère-Maxwell equation 
   \begin{equation}
       \frac{\partial \bm{B}}{\partial t}=-\nabla \times \bm{E}
   \end{equation}
   This equation is necessary, or else, when the boundary conditions $B(0)=B_0$, $B(\infty)=0$, and the continuity of the fields inside the multilayer, are imposed, there is an undetermined coefficient.

   The solutions to the equations are the following. The magnetic field is:
    \begin{equation}
        \begin{aligned}
            B_I\equiv B(0\le x < d_S)=\frac{B_0}{D}&\left\lbrace \cosh\left(\frac{x-d_S}{\lambda_1}\right)\left[\cos\left(\frac{\omega \sqrt{\varepsilon_r}}{c}d_I\right)-\frac{\omega \lambda_2\sqrt{\varepsilon_r}}{c}\sin\left(\frac{\omega \sqrt{\varepsilon_r}}{c}d_I\right)\right]\right.\\
            &\left.-\sinh\left(\frac{x-d_S}{\lambda_1}\right)\left[\frac{\lambda_2}{\lambda_1}\cos\left(\frac{\omega \sqrt{\varepsilon_r}}{c}d_I\right)+\frac{c}{\omega\lambda_1 \sqrt{\varepsilon_r}}\sin\left(\frac{\omega \sqrt{\varepsilon_r}}{c}d_I\right)\right] \right\rbrace
        \end{aligned}
    \end{equation}
    \begin{equation}
        B_{II}\equiv B(d_S\le x< d_S+d_I)=\frac{B_0}{D}\left[\cos\left(\frac{\omega\sqrt{\varepsilon_r}}{c}(x-d_S-d_I)\right)+\frac{\omega \lambda_2 \sqrt{\varepsilon_r}}{c}\sin\left(\frac{\omega\sqrt{\varepsilon_r}}{c}(x-d_S-d_I)\right)\right]
    \end{equation}
    \begin{equation}
        B_{III}\equiv B(x\ge d_S+d_I)=\frac{B_0}{D}\exp\left(-\frac{x-d_S-d_I}{\lambda_2}\right)
    \end{equation}
    Where the $D$ that appears on the denominators is defined as
    \begin{equation}
        \begin{aligned}
            D=&\cosh\left(\frac{d_S}{\lambda_1}\right)\left[\cos\left(\frac{\omega\sqrt{\varepsilon_r}}{c}d_I\right )-\frac{\omega \lambda_2\sqrt{\varepsilon_r}}{c}\sin\left(\frac{\omega \sqrt{\varepsilon_r}}{c}d_I\right)\right]+\\
            &+\sinh\left(\frac{d_S}{\lambda_1}\right)\left[\frac{\lambda_2}{\lambda_1}\cos\left(\frac{\omega\sqrt{\varepsilon_r}}{c}d_I\right )+\frac{c}{\omega \lambda_1\sqrt{\varepsilon_r}}\sin\left(\frac{\omega \sqrt{\varepsilon_r}}{c}d_I\right)\right]
        \end{aligned}
    \end{equation}

    And the electric field is
    \begin{equation}
        \begin{aligned}
            E_I=\frac{i\omega \lambda_1B_0}{D}&\left\lbrace \sinh\left(\frac{x-d_S}{\lambda_1}\right)\left[\cos\left(\frac{\omega \sqrt{\varepsilon_r}}{c}d_I\right)-\frac{\omega \lambda_2\sqrt{\varepsilon_r}}{c}\sin\left(\frac{\omega \sqrt{\varepsilon_r}}{c}d_I\right)\right]\right.\\
            &\left.-\cosh\left(\frac{x-d_S}{\lambda_1}\right)\left[\frac{\lambda_2}{\lambda_1}\cos\left(\frac{\omega \sqrt{\varepsilon_r}}{c}d_I\right)+\frac{c}{\omega\lambda_1 \sqrt{\varepsilon_r}}\sin\left(\frac{\omega \sqrt{\varepsilon_r}}{c}d_I\right)\right] \right\rbrace
        \end{aligned}
    \end{equation}
    \begin{equation}
        E_{II}=\frac{icB_0}{D\sqrt{\varepsilon_r}}\left[\sin\left(\frac{\omega\sqrt{\varepsilon_r}}{c}(x-d_S-d_I)\right)-\frac{\omega \lambda_2 \sqrt{\varepsilon_r}}{c}\cos\left(\frac{\omega\sqrt{\varepsilon_r}}{c}(x-d_S-d_I)\right)\right]
    \end{equation}
    \begin{equation}
        E_{III}=-\frac{i\omega\lambda_2 B_0}{D}\exp\left(-\frac{x-d_S-d_I}{\lambda_2}\right)
    \end{equation}
    To recover the equations in Section \ref{sec:theory}, it is assumed that $\frac{\omega}{c}\sqrt{\varepsilon_r}d_I\ll1$, since a good insulator has a relative permittivity of $\varepsilon_r \sim 10$, and working at radio-frequency implies a frequency of operation around $\omega \sim 1~\text{GHz}=10^9~\text{s}^{-1}$. This imposes a condition on the insulator layer thickness, as it must be $d_I\ll 10^{-1}~\text{m}$. This condition is held in Section \ref{sec:Num_Results}.

     \section{Calculation of the vortex equilibrium inside the first layer} \label{sec:App-Vortex_Calc}
    
    The Lorentz force that affects a vertex is proportional to the current.
    \begin{equation}
        \bm{F}=\bm{j}\times \phi_0\hat{\bm{z}},
    \end{equation}
    where $\phi_0$ is the quantum flux produced by a single vortex. In this case, two currents are at play. The first one $\bm{j}_{\text{ext}}$ is generated by the external field, and can be obtained from the Maxwell equations. If the vortex is located at $x_0$
    \begin{equation}
        \bm{j}_{\text{ext}}(x_0)=\left.\frac{1}{\mu_0}\nabla\times\bm{B}-\varepsilon_0\varepsilon_r\frac{\partial \bm{E}}{\partial t}\right|_{x=x_0}\simeq\frac{B_0}{\mu_0\lambda_1}\frac{\sinh\left(\frac{d_S-x_0}{\lambda_1}\right)+\frac{\lambda_2+d_I}{\lambda_1}\cosh\left(\frac{d_S-x_0}{\lambda_1}\right)}{\cosh\left(\frac{d_S}{\lambda_1}\right)+\frac{\lambda_2+d_I}{\lambda_1}\sinh\left(\frac{d_S}{\lambda_1}\right)}\hat{\bm{y}}.
    \end{equation}
    This current produces a force that pushes the vortex inside the superconductor layer.

    The second force arises from the interaction between the vortex and the walls of the superconductor. This can be expressed as a differential equation with boundary conditions
    \begin{equation}
    \begin{aligned}
        &-\lambda_1^2\nabla^2B+B=\phi_0\delta^{(2)}(\bm{r}-\bm{r}_0), \\
        &\partial_y B(x=0,y)=\partial_y B(x=d_S,y)=0.
    \end{aligned}
    \end{equation}
    Where $\bm{r}_0$ indicates the position of the vertex, and the Laplacian is the one in 2 dimensions $\nabla^2=\partial_x^2+\partial_y^2$. 
    We can now perform a Fourier transform on the $y$ coordinate, and since the vortex only moves along the $x$-axis, we can take $y_0=0$. Obtaining that the equation and boundary conditions have now become
    \begin{equation}
        \begin{aligned}
            \left[\partial^2_x-\left(k^2+\frac{1}{\lambda_1^2}\right)\right]B_k(x)&=-\frac{\phi_0}{\lambda_1^2}\delta(x-x_0),\\
            B_k(0)&=B_k(d_S)=0.
        \end{aligned}
    \end{equation}
    This is a Sturm-Liouville problem with a Green function, where the proposed solution can be written in terms of the solutions of the homogeneous equations and must satisfy the boundary conditions. The solution will have a discontinuity at $x_0$ due to the Green function.
    \begin{equation}
       B_k(x,x_0)=A(x_0)\sinh(px)\Theta(x-x_0)+C(x_0)\sinh(p(x-d_S))\Theta(x_0-x).
    \end{equation}
    Where $p=\sqrt{k^2+(1/\lambda^2)}$, and $\Theta(x)$ is the Heaviside Theta function. Finally, applying the continuity and jump conditions at $x=x_0$, which in this case are
    \begin{equation}
    \begin{aligned}
        A(x_0)\sinh(px_0)=C(x_0)\sinh(p(x_0-d_S)), \\
        C(x_0)\cosh(p(x_0-d_S))-A(x_0)\cosh(px_0)=-\frac{\phi_0}{p\lambda_1^2}.
    \end{aligned}
    \end{equation}
    After solving these equations, it can be seen that the magnetic field is 
    \begin{equation}
        B_k(x,x_0)=\frac{\phi_0}{2p\lambda_1^2\sinh(pd_S)}\left\lbrace\cosh[p(|x-x_0|-d_S)]-\cosh[p(x+x_0-d_S)]\right\rbrace.
    \end{equation}
    Now, the current that is applied to the vortex when this one is at $x_0$ can be computed
    \begin{equation}
        \bm{j}_{\text{vor}}(x_0,0)=\left.\frac{1}{\mu_0}\nabla\times\bm{B}\right|_{\substack{x=x_0\\ y=0}}=-\frac{1}{\mu_0}\left.\int\frac{{\rm d}k}{2\pi}\ \partial_x B_k(x)\right|_{x=x_0}\hat{\bm{y}}.
    \end{equation}
    Here, some caution is required because the vortex self-interaction is divergent. This manifests in the derivative as a discontinuity in the current.
    \begin{equation}
        \left.\partial_x B_k(x)\right|_{x=x_0+0^{\pm}}=-\frac{\phi_0}{2\lambda_1^2}\left[\pm1+\frac{\sinh[p(2x_0-d_S)]}{\sinh(pd_S)}\right].
    \end{equation}
    The $\pm1$ term is discontinuous and produces a divergence. By removing this term, the current produced by the interaction with the walls is
    \begin{equation}
        \bm{j}_{\text{vor}}(x_0)=\frac{\phi_0}{4\pi\mu_0\lambda_1^2}\int^{\infty}_{-\infty}{\rm d}k\frac{\sinh[p(2x_0-d_S)]}{\sinh(pd_S)}\hat{\bm{y}}.
    \end{equation}
    When dealing with distances much smaller than $\lambda_1$, we can approximate $p\simeq |k|$. And the integral has an analytical solution
    \begin{equation}
        \bm{j}_{\text{vor}}(x_0)=-\frac{\phi_0}{4\mu_0d_S \lambda_1^2}\cot\left(\frac{\pi x_0}{d_S}\right)\hat{\bm{y}}\stackrel{x_0\ll d_S}{\simeq}-\frac{\phi_0}{4\pi\mu_0\lambda_1^2 x_0}\hat{\bm{y}}.
    \end{equation}
    This current produces a force that prevents the vortex from entering the superconductor. Since the vortex has a size similar to the coherence length $\sim \xi$, we can take $x_0=\xi$. Therefore, when we compare both currents, we find a limit on the applied external magnetic field
    \begin{equation}
        B_0\le \frac{\phi_0}{4\pi \lambda_1 \xi_1}\frac{\cosh\left(\frac{d_S}{\lambda_1}\right)+\frac{\lambda_2+d_I}{\lambda_1}\sinh\left(\frac{d_S}{\lambda_1}\right)}{\sinh\left(\frac{d_S}{\lambda_1}\right)+\frac{\lambda_2+d_I}{\lambda_1}\cosh\left(\frac{d_S}{\lambda_1}\right)}.
    \end{equation}
    Since $x_0\ll d_S$, $x_0$ can be neglected in $\bm{j}_{\text{ext}}$ when doing the comparison.
    Obtaining the first condition imposed in Eq.~\eqref{eq:Max-field}.
    \section{Parameters used for the Numerical Results} \label{sec:Appendix_Num_results}
    For the different materials we have proposed, we are going to take their properties in the clean limit for simplicity and list them off in Tab.~\ref{Tab:Parameters}. However, for other mean free paths, where there may be a different impurity level, the superconducting gap does not change, and the coherence length and penetration depth change as
    \begin{equation}
        \frac{1}{\xi}=\frac{1}{\xi_0}+\frac{1}{l}, \quad \lambda=\lambda_0\sqrt{1+\frac{\xi_0}{l}},
    \end{equation}
    where $\xi_0$ and $\lambda_0$ are the coherence length and penetration depth, respectively, in the clean limit. For extreme type II superconductors, $\lambda\gg \xi$, even in the dirty limit, implying that the coherence length is much smaller than the mean free path. Therefore, in this limit, the quantities do not change regarding the impurities. 
    \begin{table}[h!]
        \centering
        \begin{tabular}{|c|c|}\hline
          Parameter   & Value  \\\hline
          $\omega$&1.3 GHz \\\hline
          $T$& 2 K\\\hline
          $\lambda_{\text{FeSe}}$& 200 nm \cite{AkiraIron} \\
             $\lambda_{\text{NbN}}$& 200 nm \cite{Bonin:1995qj}\\
             $\lambda_{\text{Nb}}$& 40 nm \cite{Bonin:1995qj, Keckert:2018nba}\\
             $\lambda_{\text{Nb}_3\text{Sn}}$&90 nm \cite{Pose_proceedings}\\\hline
             $\Delta_{\text{FeSe}}(T=2 \text{K})$& 2.2 meV\cite{FeSeRRR}\\
             $\Delta_{\text{NbN}}(T=2 \text{K})$& 2.6 meV \cite{KOMENOU1968335}\\
             $\Delta_{\text{Nb}}(T=2 \text{K})$& 1.5 meV \cite{AkiraIron}\\
             $\Delta_{\text{Nb}_3\text{Sn}}(T=2\text{K})$&3.8 meV \cite{Pose_proceedings}\\\hline
             $\sigma^{-1}_{0,\text{FeSe}}(T=300 \text{K})$& 500 $\mu\Omega\text{cm}$ \cite{AkiraIron, Tsurkan_2011}\\
             $\sigma^{-1}_{0,\text{NbN}}(T=300 \text{K})$& 70 $\mu\Omega\text{cm}$ \cite{Soldatenkova_2021, AkiraIron}\\
             $\sigma^{-1}_{0,\text{Nb}}(T=300 \text{K})$& 2 $\mu\Omega\text{cm}$ \cite{AkiraIron}\\
             $\sigma^{-1}_{0,\text{Nb}_3\text{Sn}}(T=300 \text{K})$& 35 $\mu\Omega\text{cm}$ \cite{Li_2017}\\\hline
             RRR$_{\text{FeSe}}$& 16.4 \cite{FeSeRRR}\\
             RRR$_{\text{NbN}}$& 100 \cite{eremeev:srf09-tuobau08}\\
             RRR$_{\text{Nb}}$&300 \cite{Bonin:1995qj}\\
             RRR$_{\text{Nb}_3\text{Sn}}$& 320 \cite{6663673}\\\hline
              $\xi_{\text{FeSe}}$&2.5 nm \cite{Lin_2021, AkiraIron}\\
             $\xi_{\text{NbN}}$&5 nm \cite{Bonin:1995qj}\\
             $\xi_{\text{Nb}_3\text{Sn}}$ & 4nm \cite{Bonin:1995qj, 2022APS..APRE07005P}\\\hline
             $B_{sh}^{(\text{Bulk,Nb})}$&180 mT \cite{AkiraIron, Bonin:1995qj}\\
              $B_{sh}^{(\text{Bulk,Nb}_3\text{Sn})}$&440 mT \cite{Pose_proceedings,Keckert:2018nba}\\\hline
        \end{tabular}
        \caption{Parameters used for the calculation of the different multilayer distributions in the clean limit $l\gg\xi_0$ so $\xi=\xi_0, \lambda=\lambda_L$. \cite{Pose_proceedings,Lin_2021, Tsurkan_2011,FeSeRRR,eremeev:srf09-tuobau08,AkiraIron,KOMENOU1968335,osti_1340208,Soldatenkova_2021,Bonin:1995qj,2019PhDT.........3S,2022APS..APRE07005P,1059389,Li_2017,6663673,Keckert:2018nba}}
        \label{Tab:Parameters}
    \end{table}   
\newpage
%%%%%%%%%%%%%%%%%%%%%%%%%%%%%%%%%%%%%%%%%%%%%%%%5

% This section is a list of funder names and grant numbers

  \printbibliography

\end{document}